\documentclass[12pt]{article}
\usepackage{amsmath,amsthm,amsfonts,amssymb,graphicx}
\usepackage{cite,faktor}
\usepackage{bbm,bigints}

\newlength{\xtrawidth}
\newlength{\xtraheight}
\numberwithin{equation}{section}
\numberwithin{table}{section}
\numberwithin{figure}{section}

\begin{document}

\begin{titlepage}
\begin{center}
\hfill NSF-KITP-13-144\\
\hfill BONN-TH-2013-12\\
\hfill UCSB Math 2013-17
\vskip 0.75in

{\Large \bf Perturbative Corrections to K\"ahler Moduli Spaces}\\

\vskip 0.4in

{ James Halverson${}^{a}$, Hans Jockers${}^{b}$, Joshua M.~Lapan${}^{c}$, David R.~Morrison${}^{d}$}\\

\vskip 0.3in
{\footnotesize
\begin{tabular}{ll}
${}^{\, a}${\em KITP,} &
${}^{\, b}${\em Bethe Center for Theoretical Physics,} \\
$\phantom{{}^{\, a}}${\em University of California} &
$\phantom{{}^{\, b}}${\em Physikalisches Institut, Universit\"at Bonn} \\
$\phantom{{}^{\, a}}${\em Santa Barbara, CA 93106, USA} \qquad\qquad\qquad &
$\phantom{{}^{\, b}}${\em 53115 Bonn, Germany} \\[4ex]
${}^{\, c}${\em Department of Physics,} &
${}^{\, d}${\em Departments of Mathematics and Physics,} \\
$\phantom{{}^{\, c}}${\em McGill University} &
$\phantom{{}^{\, d}}${\em University of California} \\
$\phantom{{}^{\, c}}${\em Montr\'eal, QC, Canada} &
$\phantom{{}^{\, d}}${\em Santa Barbara, CA 93106, USA} \\[2ex]
\end{tabular}
}
\end{center}

\vskip 0.35in

\begin{center} {\bf Abstract} \end{center}

We propose a general formula for perturbative-in-$\alpha'$ corrections to the K\"ahler potential on the quantum K\"ahler moduli space of Calabi--Yau $n$-folds, for any $n$, in their asymptotic large volume regime. The knowledge of such perturbative corrections provides  an important ingredient needed to analyze the full structure of this K\"ahler potential, including nonperturbative corrections such as the Gromov--Witten invariants of the Calabi--Yau $n$-folds. We argue that the perturbative corrections take a universal form, and we find that this form is encapsulated in a specific additive characteristic class of the Calabi--Yau $n$-fold which we call the log Gamma class, and which  arises naturally in a generalization of Mukai's modified Chern character map.  Our proposal is inspired heavily by the recent observation of an equality between the partition function of certain supersymmetric, two-dimensional gauge theories on a two-sphere, and the aforementioned K\"ahler potential. We further strengthen our proposal by comparing our findings on the quantum K\"ahler moduli space to the complex structure moduli space of the corresponding mirror Calabi--Yau geometry.

\vfill

\noindent v2: August 30, 2013

{
\let\thefootnote\relax\footnotetext{jim@kitp.ucsb.edu, jockers@uni-bonn.de, jlapan@physics.mcgill.ca, drm@math.ucsb.edu}
}

\end{titlepage}

\section{Introduction}

Two-dimensional $\mathcal{N}=(2,2)$
nonlinear $\sigma$-models with Calabi--Yau target manifolds
 have K\"ahler moduli spaces, described semiclassically as
the complexification of the K\"ahler cone of the Calabi--Yau manifold.
The metrics on these moduli spaces admit perturbative as well as 
non-perturbative corrections from their semiclassical approximations.
The perturbative corrections for Calabi--Yau threefolds were determined
by Candelas, de la Ossa, Green, and Parkes \cite{CDGP} and are related to
the four-loop $\sigma$-model contributions to the $\beta$-function
 calculated in 
\cite{Grisaru:1986dk}.

Interestingly, there is no five-loop correction to the $\beta$-function
\cite{Grisaru:1986wj}.
However, a framework for discussing perturbative corrections
to all orders was laid out in \cite{Freeman:1986zh}.
Nemeschansky and Sen \cite{Nemeschansky:1986yx}
explained how to modify the Calabi--Yau metric with nonlocal field
redefinitions to achieve vanishing $\beta$-function in spite of
loop corrections (of apparently arbitrary order).

Recently, a connection has been found between the (ultraviolet)
partition function
on $S^2$ of certain two-dimensional supersymmetric field theories
and the metric on the K\"ahler moduli spaces of the conformal theories
to which they flow under renormalization.  (The connection was discovered
in \cite{2-sphere} based on calculations from \cite{Benini:2012fk,Doroud:2012xw},
and argued to be
correct in \cite{Gomis:2012wy}.)  Nonperturbative information
about the conformal field theory, such as the values of Gromov--Witten invariants, 
can be extracted 
from this partition function, but only if one has first understood
the perturbative contribution to the moduli space metric.
That is our task here.

Nonperturbative information has been extracted in the case of 
 Calabi--Yau threefolds in 
\cite{2-sphere,Park:2012nn,Sharpe:2012fk,new-methods,Bonelli:2013rja,Sharpe:2013bwa}, 
and 
for a class of Calabi--Yau fourfolds in \cite{arXiv:1302.3760}.  
In this note,
we will propose a general formula for the perturbative contribution
 which involves the so-called ``Gamma
class'' that was first discovered in a prescient paper of Libgober 
\cite{arXiv:math.AG/9803119} and later rediscovered by Iritani
\cite{arXiv:0712.2204,MR2553377}
and Katzarkov--Kontsevich--Pantev
\cite{MR2483750}.
We will verify that our formula agrees with the known perturbative
contribution for Calabi--Yau threefolds and
is consistent with the formulas in \cite{arXiv:1302.3760};
we will also check it explicitly in some other cases.
Note that the present formulation does not rely on special geometry
(a spacetime property of Calabi--Yau threefold compactifications), and
therefore applies to $\sigma$-models on Calabi--Yau $n$-folds for arbitrary $n$.

Libgober's original observation concerned the behavior of period
integrals on the mirror Calabi--Yau varieties.  These period integrals
had been computed explicitly for Calabi--Yau
hypersurfaces in toric Fano varieties \cite{alg-geom/9603014},
and Libgober observed --- generalizing an observation of Hosono,
Klemm, Theisen and Yau \cite{Hosono:1994ax}
for Calabi--Yau threefolds ---
that if he used the period integrals to compute
the $n$-point correlation functions and their derivatives
(on a Calabi--Yau $n$-fold) then the leading behavior was captured
by a combination of Chern classes dervied from the power series expansion of
the Gamma function.  Our modification of Libgober's idea is to
work with the $S^2$-partition function of the original Calabi--Yau
variety (which contains similar information
to the period integrals of the mirror) and extract the leading
behavior in this case.

The question we are addressing is a question in closed string theory, but
the correction we are proposing has a close connection to some issues
in open string theory.  In forthcoming work\footnote{This work
has now appeared \cite{Hori:2013ika}, 
along with two closely related papers \cite{Honda:2013uca,Sugishita:2013jca}.}
 of Hori and Romo
(announced in a lecture at the University of Tokyo \cite{Hori:lecture},
among other places), a computation of the partition function on
a hemisphere for these same supersymmetric field
theories is directly related to the Gamma class.
Their work provides an alternative derivation of the results we
give here.

In section \ref{sec:MP}, we review the definition of the Gamma class and its appearance in a natural modification of Mukai's Chern character map; this is relevant to mapping a particular representation of the K\"ahler potential on the complex structure moduli space (as an integral of the holomorphic $n$-form wedged with its complex conjugate) to an analogous representation of the K\"ahler potential on the K\"ahler moduli space of the mirror manifold. In section \ref{sec:pert-NLSM}, we discuss the general form of perturbative corrections to $\beta$-functions of nonlinear $\sigma$-models as well as their connection with the perturbative corrections to the K\"ahler potential on the K\"ahler moduli space of the target space~$X$. Section \ref{sec:proposal} is the centerpoint of the paper: we formulate a general proposal for the perturbative contributions to the 
K\"ahler potential based on the Gamma class;
we then compute the perturbative part of the two-sphere partition function of a class of two-dimensional abelian gauge theories associated to hypersurface Calabi--Yau $n$-folds in products of projective spaces (equivalently, this computes the perturbative part of the K\"ahler potential on the K\"ahler moduli space);  finally, we argue that the class of examples considered here is sufficiently large that the inferred form of the K\"ahler potential holds for any Calabi--Yau $n$-fold. In section \ref{sec:mirror-connection}, we discuss how to use mirror symmetry to map the K\"ahler potential of the complex structure moduli space to the K\"ahler potential of the K\"ahler moduli space of the mirror, further supporting our proposal. We close the paper with a discussion of future directions.

\section{The Mukai pairing and the Gamma class} \label{sec:MP}

\newcommand{\ch}{\operatorname{ch}}
\newcommand{\td}{\operatorname{td}}

The periods of Calabi--Yau hypersurfaces in toric Fano varieties are
generalized hypergeometric functions of the type studied in 
\cite{MR1011353}, and as such, they have a power series expansion
near any ``large complex structure'' limit point in the compactified
moduli space \cite{alg-geom/9603014}.  As is typical for hypergeometric
functions, such series expansions involve Gamma functions. Building on work of Hosono, Klemm,
Theisen and Yau~\cite{Hosono:1994ax} in the case of Calabi--Yau threefolds,\footnote{The
significance of subleading terms in the prepotential as encoding 
topological data such as the second Chern class was first recognized
in \cite{Hosono:1994ax}.} Libgober
showed \cite{arXiv:math.AG/9803119}
that the physically relevant $n$-point correlation functions
(on a Calabi--Yau $n$-fold -- cf.~\cite{higherD}) 
and their derivatives have asymptotic
expansions controlled by certain combinations of characteristic classes
of the Calabi--Yau manifold that are closely related to the Gamma function.

\subsection{Multiplicative characteristic classes}

To explain this, we need to recall Hirzebruch's notion \cite{MR1335917}
of a multiplicative characteristic class $\widehat{Q}_X$ defined for
algebraic varieties $X$, and satisfying 
$\widehat{Q}_{X\times Y}=\widehat{Q}_X\widehat{Q}_Y$.  Such a characteristic 
class can  be constructed
out of any formal power series $Q(z)$ in
a variable $z$ with constant term $1$.\footnote{One can analogously construct an
{\em additive}\/ characteristic class out of a formal power series whose
constant term is zero.}  
Since the product $Q(z_1)Q(z_2)\cdots Q(z_k)$ is symmetric in $z_1$, $z_2$,
\dots, $z_k$, it can be written as a formal power series in the
elementary symmetric functions $\sigma_1$, $\sigma_2$, \dots, $\sigma_k$:
\[ \widehat{Q}(\sigma_1,\sigma_2,\dots) \,=\, Q(z_1)Q(z_2)\cdots \]
Evaluating $\widehat{Q}$ at the Chern classes $c_1(X), c_2(X), \dots$ of
$X$
gives the characteristic class $\widehat{Q}_X$.  This takes values in the
total cohomology of $X$, and so only involves finitely many terms
in the infinite sum  $\widehat{Q}(\vec{\sigma})$ for any given $X$.

Key examples of this construction are the Todd class $\td_X$ derived from
\[ \frac z{1-e^{-z}} \,=\, 1 + \frac12 z + \frac1{12}z^2-\frac1{720}z^4
+\frac1{30240} z^6 - \cdots,\]
 which plays an important
role in the Hirzebruch--Riemann--Roch formula, and the $\widehat{A}$-genus
$\widehat{A}_X$
derived from
\[ \frac{z/2}{\sinh(z/2)} \,=\, 1 - \frac1{24}z^2+\frac7{5760}z^4
-\frac{31}{967680}z^6 +\cdots,\]
which plays an important role in anomaly inflow \cite{Green:1996dd}
and the the study of
Ramond--Ramond charges \cite{Minasian:1997mm}.
Since only even powers of $z$ occur in the $\widehat{A}$-genus, this 
characteristic class can be rewritten in
terms of Pontryagin classes and defined for more general manifolds,
not just algebraic varieties.
  Note that 
\[ \frac z{1-e^{-z}} \,=\, e^{z/2} \frac z{e^{z/2}-e^{-z/2}}
= e^{z/2} \frac{z/2}{\sinh(z/2)} \ , \]
which implies that
\[ \td_X = e^{c_1(X)/2} \widehat{A}_X\ .\]

Libgober's computations showed that the asymptotic behavior of the
$n$-point functions and their derivatives are governed by a new
multiplicative characteristic class  based on the power series for $1/\Gamma(1+z)$.
We will follow later authors \cite{arXiv:0712.2204,MR2553377,MR2483750}
instead, and base the ``Gamma class'' $\widehat{\Gamma}_X$  on the  power series expansion of
$\Gamma(1+z)$ itself, which is
$$
\begin{aligned}
 \Gamma(1+z) &= 1-\gamma z +\left(\zeta(2)+\gamma^2\right)\frac{z^2}2
-\left(2\zeta(3)+3\zeta(2)\gamma+\gamma^3\right)\frac{z^3}6\\
&{} \quad +\left(6\zeta(4)+8\zeta(3)\gamma +6\zeta(2)\gamma^2+\gamma^4\right)\frac{z^4}{24} \\
& {} \quad -\left(24\zeta(5)+20\zeta(2)\zeta(3)+27\zeta(2)^2\gamma
+20\zeta(3)\gamma^2+10\zeta(2)\gamma^3+\gamma^5\right)\frac{z^5}{120}+\dots
\end{aligned}
$$
where $\gamma$ is Euler's constant.  
(This formula can be further simplified if one wishes by using the standard facts
$\zeta(2)=\frac{\pi^2}6, \zeta(4)=\frac{\pi^4}{90}$, and so on.)

\subsection{The Mukai pairing}

There is another motivation for the introduction of the Gamma class.
As shown long ago by Atiyah and Hirzebruch \cite{MR0139181},
the Chern character defines a ring homomorphism from topological
K-theory to cohomology
\[ \ch:
K(X) \to H^{\text{even}}(X,\mathbb{Q}) \ , 
\]
which is an isomorphism after tensoring the left
side with $\mathbb{Q}$.  
We will be primarily concerned with those elements of $K(X)$ that
can be realized as coherent sheaves on $X$; the subgroup of such
elements will be denoted $K_{hol}(X)$, following 
\cite{Hosono:2000eb}.\footnote{%
The main point of \cite{Hosono:2000eb} is the proposal that the
correct integral structure in the A-model is provided by $K_{hol}(X)$.
}

In the case of K3 surfaces,
Mukai \cite{MR893604} 
considered the natural bilinear
pairing on
$K_{hol}(X)$
given by the holomorphic Euler characteristic 
\[\chi(\mathcal{E},\mathcal{F})\,:=\, \sum_k (-1)^k 
\dim \operatorname{Ext}^k_{\mathcal{O}_X}(\mathcal{E},\mathcal{F}) \ ,\]
and explained how to construct a bilinear pairing on 
$H^{\text{even}}(X,\mathbb{Z})$ and a modified Chern character
map
\[ \mu: K_{hol}(X) \to H^{\text{even}}(X,\mathbb{Q}) \ , \]
which is an isometry, i.e., preserves the bilinear 
pairings.\footnote{In fact, in Mukai's original construction $\mu$
reversed the sign on the bilinear pairing,
but we are following
the more recent convention of preserving the sign.}

Mukai's construction is based on the Hirzebruch--Riemann--Roch formula
\cite{MR1335917}, which in this context says
\[\chi(\mathcal{E},\mathcal{F}) \,=\, \int_X \ch \mathcal{E}^\vee \wedge \ch \mathcal{F} \wedge \td_X \ ,\]
where $\td_X$ is the Todd class.  
The Todd class has a square root based on the square root power series
with constant term $1$:
\[ \sqrt{\frac z{1-e^{-z}}} = 1+\frac z4+\frac{z^2}{96}-\frac{z^3}{384}
-\frac{z^4}{10240}+\frac{19z^5}{368640}+\frac{79z^6}{61931520} -\frac{55z^7}{49545216}+\dots \ .\]
Mukai's modified Chern character
map is defined by
\[ \mu(\mathcal{E}) := \ch \mathcal{E} \, \sqrt{\td_X}\ ,\]
where we have implicitly extended the definition to the entire K-theory
group $K(X)$.
(We omit the wedge product symbol since all differential forms here are of even
degree and hence commute.)  This construction works in 
principle for any algebraic variety~$X$.%
\footnote{For Calabi--Yau threefolds the corresponding terms in periods and prepotentials
were first computed in \cite{Hosono:1994ax} with a generalization for Calabi--Yau $n$-folds
proposed in~\cite{arXiv:math.AG/9803119}. For Calabi--Yau fourfolds certain integral
periods are worked out in~\cite{Grimm:2009ef}.} 

However, as C{\u a}ld{\u a}raru 
has  explained in section~3 of \cite{MR2141853},
defining the Mukai pairing for a general algebraic variety $X$ requires
some care.  The pairing takes the general form
\[ \langle v\ |\  w \rangle = \int_X v^\vee \wedge w \]
for some ``duality'' involution $v\mapsto v^\vee$, and 
C{\u a}ld{\u a}raru shows that this involution must be
\[ v^\vee = \frac{\tau(v)}{\sqrt{\ch \omega_X}}\ ,\]
where $\omega_X=\mathcal{O}_X(K_X)$ is the canonical divisor on $X$
whose Chern character is $\ch(\omega_X)=\exp(-c_1(X))$, and where
$v\mapsto \tau(v)$ is the linear operator that acts as multiplication by
$(\sqrt{-1})^k$ on $H^k(X)$.\footnote{Note that to define
the Mukai pairing on even cohomology, we only need $v\mapsto \tau(v)$ to
act as multiplication by $(-1)^k$ on $H^{2k}(X)$; the formula in
the text is C{\u a}ld{\u a}raru's natural extension of this to
a pairing on the entire cohomology ring.}

As was observed in \cite{arXiv:0712.2204,MR2553377,MR2483750},
Mukai's modified Chern character map is not the only possibility:
we could instead define a map
\[ \mu_\Lambda(\mathcal{E}) := \ch \mathcal{E} \sqrt{\td_X} \exp (i\Lambda) \ ,\]
where $\Lambda$ satisfies  $\tau(\Lambda)=-\Lambda$.  (This includes Mukai's 
construction as the special case $\Lambda=0$.)
Let us verify that this map is  an isometry,  first checking how it
behaves on dual bundles.  We have
$$
\begin{aligned}
(\mu_\Lambda(\mathcal{E}))^\vee &= \frac{\tau(\mu_\Lambda(\mathcal{E}))}{\sqrt{\ch \omega_X}}\\
&= \frac{\tau(\ch \mathcal{E}) \tau(\sqrt{\td_X}) \tau(\exp(i\Lambda))}{\sqrt{\ch \omega_X}}\\
&= \frac{\ch \mathcal{E}^\vee \sqrt{\td_X} \sqrt{\ch \omega_X} \exp(-i\Lambda)}{\sqrt{\ch \omega_X}}\\
&= \ch \mathcal{E}^\vee \sqrt{\td_X} \exp(-i\Lambda) \ ,
\end{aligned}
$$
where we used the fact that 
\[ \tau(\td_X) = \ch(\omega_X) \td_X\ ,\]
as explained in proposition I.5.2 of \cite{MR801033}.
Therefore,
$$
\begin{aligned}
\langle \mu_\Lambda(\mathcal{E}) \ |\ \mu_\Lambda(\mathcal{F}) \rangle &=
\int (\mu_\Lambda(\mathcal{E}))^\vee \wedge \mu_\Lambda(\mathcal{F}) \\
&= \int \ch \mathcal{E}^\vee \sqrt{\td X} \exp(-i\Lambda) \ch \mathcal{F} \sqrt{\td X}
\exp(i\Lambda) \\
&= \int \ch \mathcal{E}^\vee \ch \mathcal{F} \td_X \ .
\end{aligned}
$$

The specific modification proposed by Iritani \cite{arXiv:0712.2204,MR2553377}
and Katzarkov--Kontsevich--Pantev \cite{MR2483750}, begins with a rewriting
of the power series associated to the Todd class using a familiar
identity from complex analysis:
\[ \frac z{1-e^{-z}} \,=\, e^{z/2} \frac{z/2}{\sinh(z/2)}
\,=\, e^{z/2}\, \Gamma(1+\tfrac z{2\pi i})\Gamma(1-\tfrac z{2\pi i})\  .\]
We use that factorization to define an alternative to the square root
of the Todd class:
if we write
\[
 \sqrt{\frac z{1-e^{-z}}} \exp(i\Lambda(z))
=
 e^{z/4} \, \Gamma(1+\tfrac z{2\pi i}) 
,\]
then since $z$ is real, we can solve for $\Lambda(z)$ as
\begin{align*}
 \Lambda(z) &= \operatorname{Im} \log \Gamma(1+\tfrac z{2\pi i})\\
&= \operatorname{Im}\left( -\gamma \frac z{2\pi i} + 
\sum_{n\ge2} (-1)^n \, \frac{\zeta(n)}n \! \left( \frac z{2\pi i}\right)^n \right)\\
&= \frac{\gamma z}{2\pi} + \sum_{k\ge1} (-1)^k\, \frac{\zeta(2k+1)}{2k+1}
\! \left(\frac z{2\pi}\right)^{2k+1} .
\end{align*}
This power series can be used to 
define an {\em additive}\/ characteristic class $\Lambda_X$
which we call the ``log Gamma class'' of $X$.
Note that since only odd powers of $z$ appear in the power series expansion,
$\tau(\Lambda_X)=-\Lambda_X$.  

In the Calabi--Yau case, when $c_1=0$, we have
\begin{equation} \label{eq:LambdaExp}
\Lambda_X \,=\, - \frac{\zeta(3)}{(2\pi)^3} c_3 + \frac{\zeta(5)}{(2\pi)^5}(c_5-c_2c_3) - \frac{\zeta(7)}{(2\pi)^7}(c_7-c_3c_4-c_2c_5+c_2^2c_3)+ \dots \ .
\end{equation}
Notice that if $X$ is a K3 surface then $\Lambda_X=0$ so the original
version of Mukai's proposal is unchanged.  
Notice also that for $X$ a Calabi--Yau threefold, the
modification is proportional to $\zeta(3) \chi/\pi^3$, where $\chi$ is
the topological Euler characteristic of $X$.

The ``replacement'' for the square root of the Todd class is thus
a muliplicative characteristic class which we call the
``complex Gamma class'':
\begin{equation} \label{eq:GRc}
 \widehat{\Gamma}^{\phantom{.}\mathbb C}_X = \sqrt{\td_X} \exp(i\Lambda_X) \ .
\end{equation} 
To compare this to the Gamma class,
let $\nu$ be the operator on the cohomology of $X$ which
acts on $H^k(X)$ as multiplication by $(2\pi i)^{k/2}$ (or the operator
on power series which multiplies $z$ by $2\pi i$).  Then since
\[\nu\left( \sqrt{\frac{z}{1-e^{-z}}} \exp(i\Lambda(z))\, e^{-z/4}\right)
= \nu\left(\Gamma(1+\tfrac z{2\pi i})\right) = \Gamma(1+z),\]
we see that
\begin{equation} \label{eq:GR}
 \nu\left(\widehat\Gamma^{\phantom{.}\mathbb C}_X \  \sqrt[4]{\ch(\omega_X)}\right)
 = \widehat\Gamma_X
\end{equation}
is the same Gamma class defined in the previous subsection.

\section{Perturbative nonlinear $\sigma$-model analysis} \label{sec:pert-NLSM}
\def\KD{H}  
\def\TD{D}  

In this section, we discuss perturbative effects in the $\mathcal{N}=(2,2)$
supersymmetric two-dimensional nonlinear $\sigma$-model on a Calabi--Yau
manifold $X$ of arbitrary dimension.  Among the marginal
operators in such a theory, we single out the ones corresponding to
the variation of the K\"ahler class of $X$, which are parameterized by
the space $H^{1,1}(X)$.  A metric on this space can be calculated in
terms of two-point correlation functions of the corresponding operators.

Since this theory has $\mathcal{N}=(2,2)$ supersymmetry, the moduli space
metric will be K\"ahler and can be described in terms of a K\"ahler potential
$K$.
If we pick (complexified) K\"ahler coordinates $t_1, \dots, t_s$ of $H^{1,1}(X)$ with respect to a basis of divisors $\KD_1, \dots, \KD_s$, we will argue that the exponentiated sign-reversed K\"ahler potential takes a particularly nice form
\begin{equation} \label{eq:Kgeneric}
  e^{-K(t)}\,=\, \int_X 
  \exp\left({4\pi \sum_{\ell=1}^s \operatorname{Im} t_\ell\,\KD_\ell} \right)\cup
  \sum_{k=0}^n \chi_k 
  \,+\, O(e^{2\pi i t}) \ ,
\end{equation}
for some cohomology classes $\chi_k\in H^{2 k}(X,\mathbb{R})$,
which specify the perturbative corrections, and where $O(e^{2\pi i t})$ represents
instanton corrections. We normalize things so that $\chi_0=1\in H^0(X)$ 
corresponds to the leading order term.
  
\subsection{Nonlinear $\sigma$-model action and the effective action}

Under the renormalization group the $\mathcal{N}=(2,2)$ supersymmetric, two-dimensional, nonlinear $\sigma$-model with K\"ahler target space $X$ (of complex dimension $n$), flows in the infrared to a conformal fixed point characterized by vanishing $\beta$-functions. In this work, the $\beta$-function of the target space K\"ahler form is of particular interest, which vanishes at tree level but is nonzero at one-loop:
\begin{equation} \label{eq:betaij}
   \frac{1}{\alpha'} \beta_{i\bar\jmath}\,=\, R_{i\bar\jmath} + \Delta\omega_{i\bar\jmath}(\alpha') \,=\,R_{i\bar\jmath}+ {\alpha'}^3 \frac{\zeta(3)}{48} T_{i\bar\jmath} + O({\alpha'}^5) \ .
\end{equation}   
Here $\alpha'$ is the coupling constant in the nonlinear $\sigma$-model. At leading one-loop order, the Ricci tensor $R_{i\bar\jmath}$ appears; $\Delta\omega_{i\bar\jmath}$ then comprises all higher loop corrections, which are exact in cohomology, i.e., $\Delta\omega = d\rho$ with some global one form $\rho$ on $X$ \cite{Howe:1986ys,Nemeschansky:1986yx}.\footnote{The finiteness of $\mathcal{N}=(4,4)$ supersymmetric non-linear $\sigma$-models with hyper-K\"ahler target spaces implies that $\Delta\omega_{a\bar b}(\alpha')=0$ \cite{Galperin:1985bj,Hull:1985at,AlvarezGaume:1985ww}.} The tensor $T_{i\bar\jmath}$ is the first non-vanishing subleading correction at four loops \cite{Gross:1986iv}, which has been explicitly calculated in ref.~\cite{Grisaru:1986px}.\footnote{The five-loop correction at order $O({\alpha'}^4)$ is absent \cite{Grisaru:1986wj}. Hence, the next non-vanishing correction is expected at order $O({\alpha'}^5)$.} Thus, at leading order the vanishing $\beta$-function $\beta_{i\bar\jmath}=0$ requires a Ricci-flat K\"ahler metric and hence a Calabi--Yau target space. However, this Ricci-flat Calabi--Yau target space metric gets further corrected at higher loops. 

For our purposes it is useful to adopt an effective action point of view for the target space geometry. Namely, we interpret the condition for the vanishing $\beta$-function as the Euler--Lagrange equation for the metric $g_{i\bar\jmath}$ arising from an action functional \cite{Gross:1986iv,Zanon:1986gg}. The relevant effective action $\mathcal{S}_{\rm eff}[g]$ takes the form
\begin{equation} \label{eq:EffAction}
  \mathcal{S}_{\rm eff}[g]\,=\, \int \sqrt{g} \left[ R(g) + \Delta S(\alpha',g) \right] \ ,
\end{equation}  
with the corrections $\Delta S(\alpha',R)$. The leading correction arises at fourth loop order ${\alpha'}^3$ and enjoys the expansion
$$
  \Delta S(\alpha',g) \,=\, {\alpha'}^3 S^{(4)}(g) + {\alpha'}^5 S^{(6)}(g) + \ldots \ .
$$
Here the $n$-th loop correction $S^{(n)}(g)$ is a scalar functional of the metric tensor and the Riemann tensor. A proposal for the structure of these terms has been put forward in ref.~\cite{Freeman:1986zh}.

\subsection{Weil--Petersson metric of the K\"ahler moduli space}

The critical locus of the effective action functional $\mathcal{S}_{\rm eff}[g]$ encodes the moduli space of K\"ahler metrics in the presence of perturbative loop corrections. We are particularly interested in the K\"ahler moduli space $\mathcal{M}_K$ of metric deformations, i.e., we consider the effective action \eqref{eq:EffAction} for the class of hermitian metrics for a fixed complex structure of the target space $X$. At leading order, the moduli space $\mathcal{M}_K$ yields the Weil--Petersson metric
\begin{equation}
     \sum_{k,\ell=1}^s G_{k\bar\ell}(t)\,dt_kd\bar t_{\bar\ell}\,=\,\partial\bar\partial K(t) \ , 
\end{equation}
in terms of the K\"ahler potential \cite{Candelas:1990pi}
\begin{equation} \label{eq:K1loop}
   e^{-K(t)}\,=\,  \frac{1}{n!} \int_X \bigg( \! {4\pi \sum_{\ell=1}^s \operatorname{Im} t_\ell\,\KD_\ell} \! \bigg)^n \!\! + O(\alpha'^3)  \ . 
\end{equation}
(Notice that the choice of (positive) constant multiplying this expression can
be altered by changing the K\"ahler potential without changing the K\"ahler
metric; our choice is a convenient normalization used in this paper.)
Clearly, since the effective action $\mathcal{S}_{\rm eff}[g]$ gets corrected beyond the leading contribution, the Weil--Petersson metric receives further corrections from higher loop orders. By means of mirror symmetry, for Calabi--Yau threefolds the four-loop correction has been determined to be \cite{CDGP}
\begin{equation} \label{eq:K4loop}
  e^{-K(t)}\,=\, \frac{1}{n!} \int_X \bigg( \! 4 \pi \sum_{\ell=1}^s \operatorname{Im} t_\ell\, \KD_\ell \! \bigg)^n \!\!
  + \frac{{\alpha'}^3}{(n-3)!}\int_X\bigg( \! 4 \pi \sum_{\ell=1}^s \operatorname{Im} t_\ell\, \KD_\ell \! \bigg)^{n-3}\!\! \cup\, \chi_3  + O(e^{2\pi i t}) \ ,
\end{equation}
for $n=3$, with the characteristic class
\begin{equation} \label{eq:4L}
  \chi_3 \,=\,   - 2\, \zeta(3) c_3(X) \ ,
\end{equation}
in terms of the third Chern class $c_3(X)$ and the Riemann $\zeta$-function. The appearance of the $\zeta$-value $\zeta(3)$ (of transcendental weight three) indicates its origin as a four-loop counterterm of the $\mathcal{N}=(2,2)$ supersymmetric nonlinear $\sigma$-model \cite{Grisaru:1986px}.

In general, further corrections in ${\alpha'}$ appear for Calabi--Yau target spaces of higher dimension $n>3$. They take the following form\footnote{Compared to eq.~\eqref{eq:Kgeneric}, we have included here the coupling constant $\alpha'$ of the nonlinear $\sigma$-model, which --- for ease of notation --- we set to $\alpha'=1$ in the other sections of this note.}
\begin{equation} \label{eq:ChiClasses}
  e^{-K(t)}\,=\, \int_X  \exp\!\bigg(\!{4\pi \sum_{\ell=1}^s \operatorname{Im} t_\ell\,\KD_\ell} \!\bigg)\cup \sum_{k=0}^n {\alpha'}^k \chi_k  +  O(e^{2\pi i t}) \ .
\end{equation}
In fact, the characteristic classes $\chi_k$ arise from the perturbative loop corrections at loop order $k+1$. Due to the appearance of higher curvature tensors in the corrections $\Delta\omega_{i\bar\jmath}$ of the $\beta$-function~\eqref{eq:betaij}, integrating such curvature tensors can be expressed in terms of the Chern classes of the tangent bundle of the target space~$X$. Furthermore, the loop corrections appearing in $\Delta\omega_{i\bar\jmath}$ at a given loop order $k+1$, i.e., at order ${\alpha'}^{k}$, give rise to corrections with transcendentality degree $k$, which  is a general property of loop corrections of supersymmetric two-dimensional $\sigma$-models \cite{Broadhurst:1996ur}. As a result, the cohomology classes $\chi_k$ are homogeneous elements of transcendental degree $k$ in the graded polynomial ring over all products of multiple $\zeta$-values up to transcendental weight $k$
\begin{equation} \label{eq:PClass}
   \chi_k \in H^{2k}(X,\mathbb{Q})[\zeta(m)_{m=2,\ldots,k},\ldots, \zeta(m_1,m_2)_{2\le m_1+m_2 \le k},\ldots,\zeta(1,\ldots,1)]_k \ .
\end{equation}
The transcendental weight of a multiple $\zeta$-value $\zeta(m_1,\ldots,m_a)$ is given by the sum $m_1+\ldots+m_a$, and the multiple zeta functions $\zeta(m_1,\ldots,m_a)$ generalize the Riemann zeta function according to \cite{MR1341859}
$$
  \zeta(m_1,\ldots,m_a) \,=\, \sum_{n_1>n_2>\ldots>n_a} \frac1{n_1^{m_1} \cdots n_a^{m_a}} \ .  
$$
Note that there are many non-trivial relations over $\mathbb{Q}$ among such multiple $\zeta$-values, see for instance~\cite{MR2578167}. 

The four-loop correction as specified by the characteristic class $\chi_3$ takes the universal form~\eqref{eq:4L}. By studying a sufficiently large class of higher-dimensional Calabi--Yau target spaces $X$, we will determine by universality the explicit characteristic classes $\chi_k$ for general $k>3$ in the following as well.

\section{Corrections from the partition function} \label{sec:proposal}

In the previous section, we argued that the perturbative contributions
to the metric on the K\"ahler moduli space take a universal form
represented by certain polynomials in the Chern classes at each fixed
degree.  We now wish to determine those Chern class polynomials, 
and we will do so
using the equivalence between the two-sphere partition function and
the exponentiated sign-reversed K\"ahler potential for a certain
large class of gauged linear sigma models \cite{2-sphere}.

First, let us state the result.  Given an integral basis $H_1$, \dots, $H_s$
of $H^{1,1}(X)$ with corresponding coordinates $t_1$, \dots, $t_s$
on the complexified K\"ahler moduli space, we will verify below that
the K\"ahler potential $K$
satisfies
\begin{equation} \label{the_answer}
 e^{-K} = (2\pi i)^n \int_X e^{-\sum_\ell (t_\ell - \bar{t}_\ell)
H_\ell} \left(
\faktor{\widehat{\Gamma}^{\phantom{.}\mathbb C}_X}
{ \overline{
\widehat{\Gamma}^{\phantom{.}\mathbb C}_X}}
\right)
+O(e^{2\pi it}).
\end{equation}
That is, the ratio of the complex Gamma class to its complex conjugate
(computed formally in the cohomology ring)
completely captures the perturbative contributions to the metric.

There is an alternative formulation using the map $\nu$ introduced in~\eqref{eq:GR} and the log Gamma class $\Lambda_X$:
\begin{equation} \label{alternate_answer}
  e^{-K} =  \int_{X} \exp\!\bigg(\!{4\pi\sum_\ell \operatorname{Im} t_\ell H_\ell} \!\bigg) \  \nu(e^{2i \Lambda_X}) + O(e^{2\pi it})\ ,
\end{equation}
from which we see that the log Gamma class contains all of the information we need to determine the classes $\chi_k$.

For a Calabi--Yau manifold $X$ of complex dimension $n$ with generically non-vanishing Chern classes $c_2(X)$ through $c_n(X)$, there are $p(k)-p(k-1)$ distinct degree $k$ monomials in the Chern classes, where $p(m)$ is  the number of partitions of the integer $m$. In the following, we consider smooth $n$-dimensional Calabi--Yau hypersurfaces in projective simplicial toric Fano varieties of dimension $n+1$. A subset of such examples of Calabi--Yau manifolds is given by smooth Calabi--Yau hypersufraces embedded in $\mathbb{P}^{n_1}\times\mathbb{P}^{n_2} \times \ldots \times \mathbb{P}^{n_s}$ (with $n_1 \le n_2 \le \ldots \le n_s$) of degree $(n_1+1,\ldots,n_s+1)$ and complex dimension $n=n_1+\ldots+n_s-1$. These latter examples already give rise to $p(n+1)$ Calabi--Yau geometries of complex dimension $n$.  In appendix~\ref{sec:NumEvidence} we demonstrate that, indeed, up to Calabi--Yau ninefolds this class of Calabi--Yau hypersurfaces yields a sufficiently generic set of examples. Therefore, it seems reasonable that the smooth Calabi--Yau hypersurfaces in projective simplicial toric Fano varieties will in general furnish sufficiently many examples to confirm our universal answer~\eqref{the_answer}.

\subsection{Gauged linear $\sigma$-model for toric hypersurfaces}

Let us consider a projective simplicial toric Fano variety $\mathbb{P}_\Sigma$ of dimension $n+1$ defined in terms of a complete fan $\Sigma$ in $N_\mathbb{R}$ of dimension $n+1$. The one-dimensional cones $\rho\in\Sigma(1)$ in the fan~$\Sigma$ are associated to toric divisors $\TD_\rho$. If the toric variety $\mathbb{P}_\Sigma$ has only zero-dimensional singularities, a generic section of the anti-canonical bundle $\mathcal{O}_{\mathbb{P}_\Sigma}(\sum_{\rho\in\Sigma(1)}\TD_\rho)$ describes a smooth Calabi--Yau hypersurface $X_\Sigma$ of dimension $n$. 

The gauged linear $\sigma$-model realizes the toric ambient space $\mathbb{P}_\Sigma$ via symplectic reduction with respect to the moment map 
\begin{equation}
\begin{aligned}
   \mu_\Sigma:\ & \mathbb{C}^{n+s+1} \rightarrow A_{n}(\mathbb{P}_\Sigma)\otimes \mathbb{R} \simeq H^{1,1}(\mathbb{P}_\Sigma) \simeq \mathbb{R}^s \ , \\
                         & (\Phi_1,\ldots,\Phi_{n+s+1}) \mapsto \frac12 \left( \sum_{k=1}^{n+s+1} q^{1}_k |\Phi_k|^2 ,\, \ldots ,\, \sum_{k=1}^{n+s+1} q^{s}_k |\Phi_k|^2 \right) \ .
\end{aligned}
\end{equation}     
Here $A_n(\mathbb{P}_\Sigma)$ denotes the $n$-th Chow group of $\mathbb{P}_\Sigma$. The integral charges $q^{\ell}_n$ are constrained by $0=\sum_{\rho\in\Sigma(1)} q^\ell_{\rho} v_{\rho}$ for $\ell=1,\ldots,s$ in terms of the one-dimensional integral generators $v_\rho$ of the cones $\Sigma(1)$. For a K\"ahler form $\omega \in H^{1,1}(\mathbb{P}_\Sigma)$ the toric variety $\mathbb{P}_\Sigma$ is now described by the symplectic quotient 
\begin{equation}
    \mathbb{P}_\Sigma \,\simeq\, \mu_\Sigma^{-1}(\omega)/U(1)^s \ ,
\end{equation}
with respect to the $U(1)^s$ action associated to the moment map $\mu_\Sigma$.\footnote{Further details on the method of symplectic reduction are reviewed for instance in ref.~\cite{MR1677117}.}

The abelian $\mathcal{N}=(2,2)$ gauged linear $\sigma$-model description of the Calabi--Yau hypersurface $X_\Sigma$ arises from the charged $\mathcal{N}=(2,2)$ chiral spectrum
\begin{equation} \label{eq:Charges}
\vbox{
\halign{\strut\vrule~~#~~\vrule&\hfil~~$#$~~\vrule&\hfil~~$#$~~&\hfil~~$#$~~&$\cdots$\hfil~~$#$~~\vrule\cr
\noalign{\hrule}
Fields & P\  & \Phi_1 & \Phi_2 & \Phi_{n+s+1}  \cr
\noalign{\hrule}\noalign{\hrule}\noalign{\hrule}
$U(1)_1$ & -q_0^1 & q_1^1 & q_2^1 & q_{n+s+1}^1 \cr
\ $\vdots$ &\vdots\hfil & \vdots\hfil & \vdots\hfil & \vdots\hfil \cr
$U(1)_s$ & -q_0^s & q_1^s & q_2^s & q_{n+s+1}^s \cr
\noalign{\hrule}
}} \ .
\end{equation}
The gauge charges of the $P$-field are given by
$$
  q_0^{\ell} \,=\, \sum_{k=1}^{n+s+1} q_k^\ell \ , \qquad \ell=1,\ldots,n+s+1 \ ,
$$  
in terms of the integral charges with respect to the remaining chiral fields. The gauge invariant superpotential reads
\begin{equation}
    W(P,\Phi) \,=\, P\,G(\Phi) \ .
\end{equation}    
Here $G(\Phi)$ is a generic polynomial in the chiral fields $\Phi_1, \ldots, \Phi_{n+s+1}$ such that $W(P,\Phi)$ forms a gauge invariant superpotential. The polynomial $G(\Phi)$ is readily identified with a section of the anti-canonical bundle $\mathcal{O}_{\mathbb{P}_\Sigma}(\sum_{\rho\in\Sigma(1)}\TD_\rho)$.

In the spectrum~\eqref{eq:Charges}, we have chosen the $U(1)$-gauge group factors of the rows such that $\mathbb{R}_{>0}^s\subset \mathbb{R}^s\simeq H^{1,1}(\mathbb{P}_\Sigma)=\operatorname{Image} (\mu_\Sigma)$ corresponds to the K\"ahler cone of the simplicial toric variety $\mathbb{P}_\Sigma$. (This amounts to identifying the rows of the charges~\eqref{eq:Charges} with the Mori cone vectors of the toric variety $\mathbb{P}_\Sigma$.) Then for any $\omega\in \mathbb{R}_{>0}^s$ the Calabi--Yau hypersurface 
\begin{equation}
   X_\Sigma \,\simeq\, \left\{ \, \mu^{-1}_\Sigma(\omega)/U(1)^s \subset \mathbb{C}^{n+s+1} \,\middle| \, G(\Phi)=0 \, \right\} \ ,
\end{equation}   
arises as the semiclassical vacuum manifold of the geometric phase of the described gauged linear $\sigma$-model with the spectrum \eqref{eq:Charges}. The choice of $\omega\in \mathbb{R}_{>0}^s$ --- which specifies the Fayet--Iliopoulos terms of the gauged linear $\sigma$-model --- determines the K\"ahler class of the hypersurface $X_\Sigma$.

\subsection{Partition function for toric Calabi--Yau hypersurfaces}

The two-sphere partition function of the abelian gauged linear $\sigma$-model with the spectrum \eqref{eq:Charges} is given by \cite{Benini:2012fk,Doroud:2012xw,2-sphere}\footnote{We have dropped an irrelevant prefactor in $Z_{S^2}(r,\theta)$ coming from the $R$-charges of the chiral fields~\eqref{eq:Charges}. For further details see ref.~\cite{2-sphere}.} 
\begin{equation} \label{eq:PF}
\begin{aligned}
    Z_{S^2}(r,\theta)\,=&\sum_{m_1,\ldots,m_s} \
    \int\limits_{\gamma+i\,\mathbb{R}^s}\!\!\!\!\frac{d\sigma_1}{2\pi i}\cdots \frac{d\sigma_s}{2\pi i}\ 
    (-1)^s\, e^{-\sum_{\ell=1}^s (4\pi r_\ell \sigma_\ell + i \theta_\ell m_\ell)} \\
    &\ \times
    \frac{\Gamma\left(\sum_{\ell=1}^s q^\ell_0 (\sigma_\ell+\frac{m_\ell}2) +1\right)}{\Gamma\left(-\sum_{\ell=1}^s q^\ell_0 (\sigma_\ell-\frac{m_\ell}2) \right)}
    \prod_{k=1}^{n+s+1} \frac{\Gamma\left(-\sum_{\ell=1}^s q^\ell_k (\sigma_\ell+\frac{m_\ell}2) \right)}{\Gamma\left(\sum_{\ell=1}^s q^\ell_k (\sigma_\ell-\frac{m_\ell}2) +1\right)}
    \ . 
\end{aligned}    
\end{equation}
in terms of the complexified K\"ahler class $r + i \theta \in \mathbb{R}_{>0}^s \oplus i\,\mathbb{R}^s$. The real vector $\gamma$ reads 
\begin{equation} \label{eq:Gamma}
  \gamma=-\varepsilon \,\hat\gamma \in \mathbb{R}^s \ , \qquad 
  \hat\gamma \in \left\{ \,x \in \mathbb{R}^s \, \middle| \, 0<\sum\nolimits_{\ell=1}^s q_k^\ell x_\ell \, \right\} \ ,
\end{equation}
with infinitesimal $\varepsilon>0$. This condition originates from the requirement of positive $R$-charges of the chiral fields \cite{Benini:2012fk}, which is essential in specifying the contours of integration in the partition function \eqref{eq:PF}. The partition function integral can be explicitly evaluated by multi-dimensional residue calculus along the lines of ref.~\cite{MR1631772}.

The perturbative quantum corrections to the partition function~\eqref{eq:PF} arise from the multi-dimensional residue at $(\sigma_1, \ldots, \sigma_s)=0$, the remaining poles give rise to non-perturbative  quantum corrections of order $O(e^{-r})$. Thus the perturbative contributions to the partition function --- which are the focus of this note --- are given by
\begin{equation} \label{eq:Zperti}
\begin{aligned}
    Z_{S^2}^{\rm pert}\,=\,&
    \int\limits_{\gamma+i\,\mathbb{R}^s}\!\!\!\!\frac{d\sigma_1}{2\pi i}\cdots \frac{d\sigma_s}{2\pi i}\
    (-1)^{n}\ \frac{\sum_{\ell=1}^s q^\ell_0 \sigma_\ell }{\prod_{k=1}^{n+s+1}\left(\sum_{\ell=1}^s q^\ell_k \sigma_\ell\right)}\\
    &\qquad\times e^{-\sum_{\ell=1}^s 4\pi r_\ell \sigma_\ell}\ 
    \frac{\Gamma\left(1+\sum_{\ell=1}^s q^\ell_0 \sigma_\ell \right)}{\Gamma\left(1-\sum_{\ell=1}^s q^\ell_0 \sigma_\ell \right)}
    \prod_{k=1}^{n+s+1} \frac{\Gamma\left(1-\sum_{\ell=1}^s q^\ell_k \sigma_\ell \right)}{\Gamma\left(1+\sum_{\ell=1}^s q^\ell_k \sigma_\ell\right)} \ , 
\end{aligned}    
\end{equation}
where the integral is taken over the $s$-dimensional plane $\gamma+i\,\mathbb{R}^s$ with $\gamma$ as defined in \eqref{eq:Gamma}. Carrying out the residues yields
\begin{equation} \label{eq:Zpert}
  Z_{S^2}^{\rm pert}\,=\, (2\pi i)^{n} \int_{X_\Sigma} e^{- \sum_\ell (\xi_\ell-\bar \xi_\ell)H_\ell} 
      \frac{\Gamma\left(1-\frac{1}{2\pi i}\sum_\rho \TD_\rho \right)}{\prod_\rho\Gamma\left(1-\frac{1}{2\pi i} \TD_\rho\right) }
      \cdot \frac{\prod_{\rho}\Gamma\left(1+\frac{1}{2\pi i} \TD_\rho \right)}{\Gamma\left(1+\frac{1}{2\pi i}\sum_\rho \TD_\rho\right)} \ ,
\end{equation}
in terms of the toric divisors $D_\rho$, $\rho=1,\ldots,n+s+1,$ and the generators $H_\ell$, $\ell=1,\ldots,s,$ of the K\"ahler cone, which are linearly-equivalent to the toric divisors according to
$$
  D_\rho \,\sim\, \sum_{\ell=1}^s q_\rho^\ell H_\ell \ .
$$
The complexified algebraic K\"ahler parameters
\begin{equation} \label{eq:Kcoord}
  \xi_\ell \,=\, -\theta_\ell + i\,r_\ell \ , \qquad  t_\ell \,=\, \xi_\ell + O(e^{-r}) \ .
\end{equation}
correspond to the K\"ahler coordinates~$t^\ell$ up to non-perturbative corrections.

The technical details of the derivation of the general expression~\eqref{eq:Zpert} are deferred to ref.~\cite{WProg}. Here, we briefly present the derivation for a hypersurface Calabi--Yau $n$-fold $X$ in a product of projective spaces 
$$
  \mathbb{P}^{n+1}_\otimes:=\mathbb{P}^{n_1}\times\mathbb{P}^{n_2} \times \ldots \times \mathbb{P}^{n_s} \ , \qquad \dim  \mathbb{P}^{n+1}_\otimes = n_1+\ldots+n_s = n+1 \ 
$$ 
(as argued before, this already furnishes a sufficient class of examples for our universality argument). The toric divisors $\TD_{\rho_\ell}$, $\rho_\ell=1,\ldots,n_\ell+1$, associated to the factors $\mathbb{P}^{n_\ell}$ in $\mathbb{P}^{n+1}_\otimes$, are linearly equivalent to the hyperplane classes $\KD_\ell$ of $\mathbb{P}^{n_\ell}$, while the Calabi--Yau hypersurface is a section of the anti-canonical bundle $\mathcal{O}_{\mathbb{P}^{n+1}_\otimes}( (n_1+1) \KD_1 + \ldots + (n_s+1) \KD_s )$. As a consequence, all non-vanishing intersection numbers of such a Calabi--Yau hypersurface $X$ read
$$
  n_{\ell} +1 \,=\, \int_X \KD_1^{n_1} \cup  \KD_2^{n_2} \cup \ldots  \cup \KD_\ell^{n_\ell -1} \cup \ldots \cup \KD_{s-1}^{n_{s-1}}\cup H_s^{n_s}  \ , \qquad \ell=1,\ldots,s \ ,
$$  
which allows us to rewrite any integral over the Calabi--Yau $n$-fold $X$ of a (formal) power series $p(\KD_\ell)$ in the hyperplane classes $\KD_\ell$ as
\begin{equation} \label{eq:identi}
\begin{aligned}  
  \int_X p(\KD_\ell) \,&=\, \left.\left( \sum_{\ell=1}^s \frac{n_\ell+1}{(n_1)! \ldots (n_\ell-1)!\ldots (n_s)!} \,\frac{d^{n}}{d\KD_1^{n_1}\cdots d\KD_\ell^{n_\ell-1}\cdots d\KD_s^{n_s}} \right) 
  p(\KD_\ell) \right|_{\KD_\ell=0} \\
  \,&=\,  \oint\limits_{0}\frac{d\sigma_1}{2\pi i}\cdots \oint\limits_{0} \frac{d\sigma_s}{2\pi i}\
     \frac{(n_1+1)\sigma_1+\ldots+(n_s+1)\sigma_s}{\sigma_1^{n_1+1}\cdots \sigma_s^{n_s+1}} \, p(\sigma_\ell) \ .
\end{aligned}  
\end{equation}
Furthermore, for the Calabi--Yau hypersurface $X$ in $\mathbb{P}^{n+1}_\otimes$, the perturbative partition function \eqref{eq:Zperti} simplifies to 
$$
\begin{aligned}
    Z_{S^2}^{\rm pert}\,=\,&
    \oint\limits_{0}\frac{d\sigma_1}{2\pi i}\cdots \oint\limits_{0} \frac{d\sigma_s}{2\pi i}\
    (-1)^{n}\ \frac{(n_1+1)\sigma_1+\ldots+(n_s+1)\sigma_s}{\sigma_1^{n_1+1}\cdots \sigma_s^{n_s+1}}\\
    &\qquad\times e^{-\sum_{\ell=1}^s 4\pi r^\ell \sigma_\ell}\ 
    \frac{\Gamma\left(1+\sum_{\ell=1}^s (n_\ell+1) \sigma_\ell \right)}{\Gamma\left(1-\sum_{\ell=1}^s (n_\ell+1) \sigma_\ell \right)}
    \prod_{\ell=1}^{s} \frac{\Gamma\left(1- \sigma_\ell \right)^{n_\ell+1}}{\Gamma\left(1+ \sigma_\ell\right)^{n_\ell+1}} \ ,
\end{aligned}    
$$
and we observe that the multi-dimensional residue calculus realizes the identity \eqref{eq:identi}. Therefore, we find for such a Calabi--Yau hypersurface $X$ in a product of projective spaces~$\mathbb{P}^{n+1}_\otimes$ agreement with the general expression~\eqref{eq:Zpert}.

Let us now make contact with the characteristic classes discussed in section~\ref{sec:MP}. Since the complex Gamma class $\widehat\Gamma^{\phantom{.}\mathbb C}$ is multiplicative, the $\widehat\Gamma^{\phantom{.}\mathbb C}$-class of the Calabi--Yau hypersurface $X_\Sigma$ embedded in $\mathbb{P}_\Sigma$ becomes
\begin{equation} \label{eq:Adj}
  \widehat\Gamma^{\phantom{.}\mathbb C}_{X_\Sigma} \,=\, \frac{\widehat\Gamma^{\phantom{.}\mathbb C}_{\mathbb{P}_\Sigma}}{\widehat\Gamma^{\phantom{.}\mathbb C}_{N_{X_\Sigma}}} \,=\, 
  \frac{\prod_{\rho} \Gamma(1+\frac1{2\pi i}\TD_\rho)}{\Gamma(1+\frac1{2\pi i}\sum_{\rho} \TD_\rho)}   \ ,
\end{equation}
where the last equality follows from the adjunction formula $N_{X_\Sigma} \simeq \mathcal{O}_{\mathbb{P}_\Sigma}(\sum_\rho \TD_\rho)$ for the normal bundle $N_{X_\Sigma}$ of $X_\Sigma$ and from the identity $c(\mathbb{P}_\Sigma) = c(\oplus_\rho \TD_\rho)$ of the total Chern classes \cite{MR1234037}. As a consequence, inserting \eqref{eq:Adj} into \eqref{eq:Zpert}, we obtain the perturbative part of the $S^2$ partition function expressed as 
\begin{equation}
  Z_{S^2}^{\rm pert}\,=\, (2\pi i)^{n} \int_{X} e^{-\sum_\ell (t_\ell -\bar t_\ell) \KD_\ell} \ \left(\faktor{\widehat{\Gamma}^{\phantom{.}\mathbb C}_X }{\overline{\widehat{\Gamma}^{\phantom{.}\mathbb C}_X }}\right) \ ,
\end{equation}
which, thanks to \cite{2-sphere},
 verifies our claim about the form of the metric for a large class
of Calabi--Yau $n$-folds.

\subsection{Further checks}

We have argued, based on the universal nature of the answer, and
a verification in a sufficient number of examples, that the metric
on the K\"ahler moduli space satisfies
\begin{equation} \label{alternate_answer-bis}
  e^{-K} =  \int_{X} e^{4\pi\sum_\ell \operatorname{Im} t_\ell H_\ell} \,\nu(e^{2i \Lambda_X}) + O(e^{2\pi it})\ .
\end{equation}
Upon expanding the exponentiated $\Lambda_X$-class, this formula makes direct contact with the perturbative cohomology classes $\chi_k$ in \eqref{eq:ChiClasses}. In particular, inserting the expansion~\eqref{eq:LambdaExp} into \eqref{alternate_answer-bis} predicts the first few perturbative corrections 
\begin{equation} \label{eq:first-few}
\begin{aligned}
   \chi_3\,&=\,-2 \zeta(3) c_3 \ , \quad \chi_4\,=\,0 \ , \quad \chi_5\,=\,2 \zeta(5) \left( c_2 c_3 -  c_5 \right) \ ,  \quad
   \chi_6\,=\, 2 \zeta(3)^2 c_3^2 \ , \\ \chi_7\,&=\, -2 \zeta(7) \left(c_2^2 c_3 -c_3c_4-c_2c_5+c_7\right) \ , \quad
\chi_8\,=\, 4\zeta(3)\zeta(5)\left( c_3c_5-c_2c_3^2\right) \ , \quad  \\
\chi_9\,&=\, -\frac43  \zeta(3)^3 c_3^3+ 2\zeta(9) \left(
 c_2^3 c_3 
{-} \frac13 c_3^3  {-} 2 c_2 c_3 c_4 
{+}  c_2^2 c_5  {+}  c_4 c_5 
{+}  c_3 c_6
{+}  c_2 c_7 
{-}  c_9 
\right) \ ,
\end{aligned}
\end{equation}  
in terms of the Chern classes $c_k$ of the Calabi--Yau $n$-fold $X$.
The contributions $\chi_3$ and $\chi_4$ are exactly what appear in
\cite{arXiv:1302.3760}.
We will verify these contributions up through $\chi_9$ 
explicitly in appendix~\ref{sec:NumEvidence}.

At four-loops, the result for $\chi_3$ agrees with the four-loop correction~\eqref{eq:4L} predicted by means of mirror symmetry in ref.~\cite{CDGP}. Furthermore, $\chi_4=0$ confirms the absence of the five-loop correction as has been established in ref.~\cite{Grisaru:1986wj}. In general the perturbative corrections $\chi_k$ are in accord with the predicted ring structure~\eqref{eq:PClass}. We futher observe that --- due to the structure of the $\Lambda_X$ class --- only ordinary $\zeta$-values occur. The absence of multiple $\zeta$-values is a bit surprising from the two-dimensional $\mathcal{N}=(2,2)$ non-linear $\sigma$-model point of view. From a mirror symmetry perspective this feature has previously been observed and explained in ref.~\cite{MR1873769}, where it is demonstrated that multiple $\zeta$-values always appear in such combinations that they can be traded by ordinary $\zeta$-values due to non-trivial transcendental relations among them.

\section{Motivation from mirror symmetry} \label{sec:mirror-connection}

Now we outline a derivation of eq.~\eqref{the_answer} from mirror symmetry to further support our proposal.
Let us begin with the mirror $Y$ of our desired Calabi--Yau manifold $X$,
and study a large complex structure limit point in the complex structure
moduli space of $Y$.  There are various assumptions we could make
about the local structure near such a point (analyzed thoroughly in \cite{compact}), but let us take the simplest form: we assume that our (complex
structure) moduli space 
$\mathcal{M}_0$ is
compactified to $\mathcal{M}$
so that the boundary 
$\partial \mathcal{M} = \mathcal{M} - \mathcal{M}_0$
is a divisor with simple normal
crossings, and so that our large complex structure limit point is
a point of maximal depth in the boundary.  That is, we assume
there are local coordinates $z_1$, \dots, $z_s$ on the compactified moduli
space $\mathcal{M}$
such that our point is the origin, and such that the boundary divisor has
components $\{z_\ell=0\}$.  Our task is to study the asymptotic behavior
of the so-called Weil--Petersson metric on the moduli space.

The basic structure of the metric was found long ago by Tian~\cite{Tian}
and Todorov~\cite{Todorov-weil-petersson}, who showed
that if $\Omega_z$ is a family of holomorphic $n$-forms depending
on the parameter~$z$, then the K\"ahler potential $K$ of the Weil--Petersson
metric (with a suitable normalization) satisfies
\[ e^{-K} \,=\, (-1)^{\frac{n(n-1)}2}\,(2\pi i)^n\int_Y \Omega_z \wedge \overline{\Omega_z}\ ,\]
where, as elsewhere in this paper, $n$ is the dimension of the Calabi--Yau manifold $Y$.
This is the quantity we wish to compare to our results about the
metric on the K\"ahler moduli space of the mirror.

Let $Y_z$ be the Calabi--Yau manifold associated to $z$, for $z$ not in
the boundary.  Then the cohomology groups $H^n(Y_z,\mathbb{Z})$ do not
form a single-valued $\mathbb{Z}$-bundle, but rather  undergo a
monodromy transformation $T_\ell$ as we loop around
$\{z_\ell=0\}$ --- such a structure is known as a {\em local system}.
Let $N_\ell=\log T_\ell$.
We can find some single-valued objects defined throughout our coordinate
patch of the form
\[  \exp\!\Big(-\sum_\ell \frac1{2\pi i} N_\ell \log z_\ell\Big)\,v\]
for any $v\in H^n(Y_z,\mathbb{Z})$.  The point is that as the argument
of $z_\ell$ increases through $2\pi$, following $v$ continuously takes it
to $T_\ell\,v$, but this is then canceled by the factor of $\exp(-N_\ell)=T_\ell^{-1}$ we inserted into the expression.

These single-valued objects capture the possible asymptotic behaviors of
Hodge-theoretic quantities near the large complex structure limit point.
In particular, the asymptotic behavior of a family $\Omega_z$ of
holomorphic $n$-forms, which depend holomorphically on $z$, must take
the form
\begin{equation} \label{eq:asymptotic}
  \exp\!\Big(-\sum_\ell \frac1{2\pi i} N_\ell 
\log z_\ell\Big)\,v_{\text{holo}} + O(z) 
\end{equation}
for some appropriate vector $v_{\text{holo}}\in H^n(Y_z,\mathbb{C})$.
We use complex coefficients for cohomology here because the holomorphic form
can be modified by an arbitrary constant.

When written in this form, we can calculate the asymptotic behavior of
the Tian--Todorov exponentiated sign-reversed
K\"ahler 
potential $e^{-K}=(-1)^{\frac{n(n-1)}2}(2\pi i)^n
\int 
\Omega_z\wedge\overline{\Omega_z}$.  It is
\begin{align*}
&
 (-1)^{\frac{n(n-1)}2}(2\pi i)^n
\bigintssss_Y \exp\!\Big(-\sum_\ell \frac1{2\pi i} N_\ell \log z_\ell\Big)\,v_{\text{holo}}
\wedge
\exp\!\Big(\sum_\ell \frac1{2\pi i} N_\ell \log \overline{z_\ell}\Big)\,
\overline{v_{\text{holo}}}
\\
&
\sim
 (-1)^{\frac{n(n-1)}2}(2\pi i)^n
\bigintssss_Y \exp\!\Big(-\sum_\ell \frac1{2\pi i} N_\ell (\log z_\ell
+\log \overline{z_\ell}) \Big)\,v_{\text{holo}}
\wedge
\overline{v_{\text{holo}}} \ ,
\end{align*}
where $\sim$ means that the two sides differ by $O(z)$.
If we rewrite this in terms of  $t_\ell=\frac1{2\pi i}\log z_\ell$,
the asymptotic expression becomes
\begin{equation} \label{eq:rewritten}
\begin{aligned} e^{-K}
&\sim 
 (-1)^{\frac{n(n-1)}2}(2\pi i)^n
\bigintssss_Y \exp\!\Big(-\sum_\ell N_\ell (t_\ell-\overline{t_\ell})
 \Big)\,v_{\text{holo}}
\wedge
\overline{v_{\text{holo}}}
\\
&\sim  
\sum_k\frac1{k!}
\bigintssss_Y
\Big(4\pi \sum_\ell  N_\ell \, \operatorname{Im} t_\ell   \Big)^k
v_{\text{holo}}
\wedge
(-1)^{\frac{n(n-1)}2}
(2\pi i)^{n-k}\ \overline{v_{\text{holo}}} \ , 
\end{aligned}
\end{equation}
which we can recognize as being the same general form as \eqref{eq:Kgeneric},
once a specific choice for $v_{\text{holo}}$ has been made.

\subsection{The Gauss--Manin connection and the Hodge filtration}

The family of holomorphic $n$-forms $\Omega_z$ can be regarded as a
section of a holomorphic bundle $\mathcal{H}$ over the moduli space
whose fibers are the cohomology groups $H^n(Y_z,\mathbb{C})$.
Sections of $\mathcal{H}$ can be differentiated with respect
to parameters using the
{\em Gauss--Manin connection}\/
$\nabla: \mathcal{H} \to \mathcal{H}\otimes \Omega^1_{\mathcal{M}_0}$,
which has the property that $\nabla(\varphi)=0$ if $\phi_z\in H^n(Y_z,\mathbb{Z})$ for all $z$.  The holomorphic $n$-forms span a rank one subbundle
$\mathcal{F}^n$ of $\mathcal{H}$ which is the first subbundle in the
{\em Hodge filtration}:
\[ \mathcal{F}^p \ = \ \big\{\phi \ \big| \ \phi_z \in H^{p,n-p}(Y_z,\mathbb{C}) \oplus H^{p+1,n-p-1}(Y_z,\mathbb{C}) \oplus \cdots \oplus H^{n,0}(Y_z,\mathbb{C}) \ \big\} \ \subset \
\mathcal{H} \ .\]
We can think of elements of $\mathcal{F}^p$ as consisting of sums of forms in $H^n(Y,\mathbb{C})$, each of which has at least $p$ holomorphic differentials. The subbundles in this filtration are related by differentiation:
\[ \nabla(\mathcal{F}^p)\subset\mathcal{F}^{p-1}\otimes \Omega^1_{\mathcal{M}_0}.\]

As mentioned above, the integer cohomology groups 
(contained in $\{\varphi \ |\ \nabla(\varphi)=0\}$) form a local
system with monodromy around the boundary divisors; any explicit
description of this local system must  involve
logarithms of the coordinates.  Since the derivative of a logarithm has
a simple pole, it is natural to expect (and known to be true
\cite{MR0417174}) that the bundle ${\mathcal H}$
and the Gauss--Manin connection $\nabla$
extend over
 $\mathcal{M}$ to
a bundle $\widetilde{\mathcal{H}}$ and a connection $\widetilde{\nabla}$
satisfying
\[ \widetilde{\nabla}: \widetilde{\mathcal{H}}
\to \widetilde{\mathcal{H}}\otimes \Omega^1_{\mathcal{M}}(\log 
\partial \mathcal{M}),\]
where $\Omega^1_{\mathcal{M}}(\log
\partial \mathcal{M})$ is the sheaf of differentials with logarithmic
poles on the boundary (generated in our local coordinate system by
$dz_1/z_1$, \dots, $dz_s/z_s$).
It is also known
 \cite{MR0382272} that the Hodge filtration extends to a filtration
$\widetilde{\mathcal{F}}^p$ by holomorphic subbundles satisfying
\[ \widetilde{\nabla}(\widetilde{\mathcal{F}}^p) \subset 
\widetilde{\mathcal{F}}^{p-1} \otimes \Omega^1_{\mathcal{M}}(\log
\partial \mathcal{M}).\]

\subsection{The mirror}

The mirror analogues of the  local system, the 
extended Gauss--Manin
connection, and the extended Hodge filtration were described in 
\cite{parkcity}
 (see also \cite{delignelimit}.)
These provide the mirror analogue of the
family of holomorphic $n$-forms, which can then be used to calculate the
metric on the K\"ahler moduli space.  
However, one ingredient of this computation (the complex conjugation of the
Hodge bundles) was left implicit in \cite{parkcity}, and in fact needs 
to be modified as we shall describe.

The mirror of the logarithms $N_\ell$ of the monodromy transformations,
which define the local system,
are the transformations on $H^{\text{even}}(X,\mathbb{C})$ given
by cup-product with 
the
corresponding divisors $\KD_\ell$, which are the
edges of the K\"ahler cone.\footnote{More precisely, we have chosen
a simplicial cone within the K\"ahler cone, and the $\KD_\ell$ are the
edges of it.}
The  vector that plays the r\^ole of $v_{\text{holo}}$ in
\eqref{eq:asymptotic}
is the generator
$\mathbbm{1}\in
H^{0,0}(X)$, so that a generator of the mirror of $\widetilde{\mathcal{F}}^n$
is
\[
  \exp\!\Big(-\sum_\ell \frac1{2\pi i} N_\ell 
\log z_\ell\Big)\,\mathbbm{1} + O(z). 
\]

The local system $\bigcup H^n(Y_z,\mathbb{Z})$
has an integer structure, and the earliest guesses about
that structure on the mirror were to use the integer structure
on even cohomology $H^{\text{even}}(X_z,\mathbb{Z})$.
However, as Hosono pointed out \cite{Hosono:2000eb}, the correct
integer structure --- the one which is compatible with mirror symmetry
in known examples --- 
 is provided by $K_{hol}(X)$.
Thus, whenever we need the integer structure of the local system
$\bigcup H^{\text{even}}(X_z,\mathbb{C})$, we should apply the
isomorphism between $K_{hol}(X)\otimes \mathbb{C}$ and 
$ H^{\text{even}}(X,\mathbb{C})$ and use the integer structure
coming from $K$-theory.

For the purpose of calculating the metric, we do not need the integer
structure on the local system, but we do need the real structure
so that we can perform complex conjugation.\footnote{In \cite{parkcity},
the complex conjugation was implicitly assumed to come from the
integer structure on cohomology; we are correcting that assumption here.}
And as 
Iritani \cite{arXiv:0712.2204,MR2553377} and Katzarkov--Kontsevich--Pantev
\cite{MR2483750} have taught us, the most natural isomorphism to use
is not the Chern character map, but 
rather
\begin{equation} \label{eq:isom}
\mu_{\Lambda_X}: \mathcal{E} \mapsto \ch(\mathcal{E}) \wedge 
{\widehat{\Gamma}^{\phantom{.}\mathbb C}_X}.
\end{equation}
Note that neither the Gauss--Manin connection, nor the Hodge filtration, 
nor even the
local system with complex coefficients --- the topics 
discussed in \cite{parkcity} --- depend on this isomorphism.  But the complex
conjugation needed to define the metric does depend on it.

To compute the effect of complex conjugation, note that wedging
with $\widehat{\Gamma}^{\phantom{.}\mathbb C}_X$ is
an invertible operation, since the power series for 
$\Gamma(1+\frac z{2\pi i})$ is invertible.  Thus, the complex conjugate
of $w$ is computed using \eqref{eq:isom} as
\[ w \mapsto \ch^{-1}(\faktor w{\widehat{\Gamma}^{\phantom{.}\mathbb C}_X})
\mapsto \ch^{-1}(\faktor{\overline{w}}{\overline{\widehat{\Gamma}^{\phantom{.}\mathbb C}_X}})
\mapsto \overline{w} \left( \faktor{\widehat{\Gamma}^{\phantom{.}\mathbb C}_X}{
\overline{\widehat{\Gamma}^{\phantom{.}\mathbb C}_X}}\right).\]

Finally, the mirror of the wedge product pairing 
$\langle \varphi \ |\ \psi \rangle := (-1)^{n(n-1)/2} \int_Y 
\varphi \wedge \psi$ is the Mukai pairing
$\langle v \ |\ w \rangle := \int_X v^\vee \wedge w$.  Thus, the
asymptotic behavior of the metric, by a computation parallel to
\eqref{eq:rewritten}, is seen to be
\begin{align*}
&
 (2\pi i)^n
\bigintssss_X \exp\!\Big(-\sum_\ell H_\ell (t_\ell-\overline{t_\ell})
 \Big)\,\mathbbm{1}
\wedge
\overline{\mathbbm{1}} 
\left( \faktor{\widehat{\Gamma}^{\phantom{.}\mathbb C}_X}{
\overline{\widehat{\Gamma}^{\phantom{.}\mathbb C}_X}}\right) 
\\
& \sim \, (2\pi i)^n
\bigintssss_X \exp\!\Big(-\sum_\ell H_\ell (t_\ell-\overline{t_\ell})
 \Big)
\wedge
\left( \faktor{\widehat{\Gamma}^{\phantom{.}\mathbb C}_X}{
\overline{\widehat{\Gamma}^{\phantom{.}\mathbb C}_X}}\right) ,
\end{align*}
which agrees with our previous proposal.

\section{Discussion}

Using the recently-proposed two-sphere partition function
correspondence \cite{2-sphere}, we determined the perturbative
$\alpha'$-corrections to the K\"ahler potential for the
Weil--Petersson metric of the quantum K\"ahler moduli space of
Calabi--Yau $n$-folds $X$. From the associated $\mathcal{N}=(2,2)$
nonlinear $\sigma$-model perspective we derived the perturbative
corrections to the marginal K\"ahler deformations of their target
spaces. As we explained, these quantum corrections appear from the
requirement of a vanishing $\beta$-function for the K\"ahler metric
$g_{i\bar\jmath}$ of the target space manifold $X$. To leading order
--- i.e., to one-loop order --- this condition restricts the target
space metric to the Ricci-flat Calabi--Yau metric in a given K\"ahler
class \cite{MR480350}. The subleading higher-loop corrections further
modify this target space metric (within its original K\"ahler
class) \cite{Howe:1986ys,Nemeschansky:1986yx}. Thus --- although not
spelled out in detail here --- our calculation fixes coefficients of
the counterterms proposed in~\cite{Freeman:1986zh}, which come
from higher-loop perturbative corrections to the $\beta$-function of
the K\"ahler metric to the Calabi--Yau target space~$X$. In
particular, our proposal is in accord with the four-loop correction
predicted by means of mirror symmetry in~\cite{CDGP}, and it
confirms the absence of a five-loop correction established
in~\cite{Nemeschansky:1986yx}. Furthermore, our proposed perturbative
corrections are in agreement with the quantum K\"ahler potential for
Calabi--Yau fourfolds conjectured in~\cite{arXiv:1302.3760}.

We found that the structure of the perturbative $\alpha'$-corrections
discussed above is captured by interesting characteristic classes of
the Calabi--Yau $n$-fold~$X$. The perturbative corrections were most
conveniently extracted from an additive characteristic class  which we introduced, the ``log
Gamma'' class $\Lambda_X$, which in turns is
closely related to the multiplicative Gamma class $\widehat\Gamma_X$
introduced
in~\cite{arXiv:math.AG/9803119,arXiv:0712.2204,MR2553377,MR2483750}. We
explained how the ``log Gamma'' class~$\Lambda_X$ is part of a natural
generalization to Mukai's modified Chern character
map. Previously, the characteristic class $\widehat\Gamma_X$ had
appeared in period integrals on mirror Calabi--Yau
geometries \cite{arXiv:math.AG/9803119}\footnote{As well as 
quantum cohomology of Fano varieties \cite{math.AG/9807034}.} 
and their generalizations
\cite{arXiv:0712.2204,MR2553377,MR2483750}, and also in the context of
deformation quantization of Poisson manifolds
\cite{MR1718044,MR2483750}. It is gratifying to see that our
two-sphere partition function calculation --- which is
mirror-symmetric to the period integral analysis of
\cite{arXiv:math.AG/9803119} --- conforms with these other approaches.

The ``log Gamma'' characteristic class $\Lambda_X$ also has
interesting number theoretic properties due to the appearance of
Riemann $\zeta$-values. In particular, the form-degrees of the terms
in the characteristic class $\Lambda_X$ conform with the
transcendentality degrees of the $\zeta$-values which appear
there. From a physics point of view this property indicated that the
terms in $\Lambda_X$ originated from perturbative corrections of
two-dimensional supersymmetric field theories
\cite{Broadhurst:1996ur}, where both the form degree and the
transcendentality degree signal the loop order of the corresponding
perturbative correction to the aforementioned $\beta$-function of the
target space metric of $X$.

Finally, let us point out that the perturbative corrections to the
K\"ahler potential of the quantum K\"ahler moduli space which we have
determined, together with the generalization of Mukai's modified Chern
character map which we have studied, provide the necessary ingredients
to define a variation of polarized Hodge structures on the
closed-string topological A-periods in the asymptotic large volume
limit \cite{parkcity,MR2483750}. By deforming the ordinary cohomology
ring to the A-model quantum cohomology ring this asymptotic variation
of polarized Hodge structures canonically extends to the variation of
polarized Hodge structures of topological A-periods beyond the
asymptotic large volume limit. In this way, we can systematically
derive the exponentiated sign-reversed K\"ahler potential for any
Calabi--Yau $n$-fold --- including both perturbative and
non-perturbative corrections --- from first principles. In particular,
as proposed for Calabi--Yau threefolds in~\cite{2-sphere} and
Calabi--Yau fourfolds in~\cite{arXiv:1302.3760}, this allows us to extract certain
Gromov--Witten invariants of Calabi--Yau $n$-folds, for any $n$, from
the sphere-partition function proposal \cite{2-sphere}. We will
discuss such implications of this work in detail in ref.~\cite{WProg}.

\bigskip
\subsection*{Acknowledgments}

We would like to thank Vijay Kumar and Mauricio Romo for participation
in the early stages of this project, the Graduate School of Mathematical
Sciences, University of Tokyo, for hospitality at the inception of
the project, and Kentaro Hori for sharing insights about his
work in progress \cite{Hori:lecture}.  We also thank
Ruth Britto, Eduardo Cattani, Albrecht Klemm, Shinobu Hosono, Peter Mayr, Jan Manschot, Greg Moore, Stephan Stieberger and Stefan Theisen
for useful discussions and correspondence.
D.R.M. thanks 
the Kavli Institute for the Physics and Mathematics of the Universe, 
the Aspen Center for Physics, and the Simons Center for Geometry and Physics
for hospitality; he also thanks the organizers of Strings 2013 for the
opportunity to present these results there.
J.H.~is supported in part by the National Science Foundation under 
Grant No.\ PHY11-25915;
H.J.~is supported by the DFG grant KL 2271/1-1;
J.M.L.~is supported by the National Science and Engineering Research Council of Canada;
D.R.M.~is supported in part by National Science Foundation Grants DMS-1007414 and PHY-1066293.

\bigskip
\appendix

\section{Numerical evidence} \label{sec:NumEvidence}

\begin{table}[t]
  \centering
  \begin{tabular}{|c|c|}
  \hline
    \strut Toric Ambient Space $\mathbb{P}^{n+1}_\otimes$ & Perturbative Correction $\chi_n$ \\ \hline\hline
$\mathbb{P}^{4} $ & $400\,\,\zeta(3)$ \\ \hline
$\mathbb{P}^{5} $ & $0$ \\
$\mathbb{P}^{4} \times \mathbb{P}^{1} $ & $0$ \\ \hline
$\mathbb{P}^{6} $ & $47040\,\, \zeta(5)$ \\
$\mathbb{P}^{5} \times \mathbb{P}^{1} $ & $37320\,\, \zeta(5)$ \\ \hline
$\mathbb{P}^{7} $ & $451584\,\, \zeta(3)^2$ \\
$\mathbb{P}^{6} \times \mathbb{P}^{1} $ & $357504\,\, \zeta(3)^2$ \\
$\mathbb{P}^{5} \times \mathbb{P}^{2} $ & $321408\,\, \zeta(3)^2$ \\
$\mathbb{P}^{5} \times \mathbb{P}^{1} \times \mathbb{P}^{1} $ & $285696\,\, \zeta(3)^2$ \\ \hline
$\mathbb{P}^{8} $ & $12299040\,\, \zeta(7)$ \\
$\mathbb{P}^{7} \times \mathbb{P}^{1} $ & $9586976\,\, \zeta(7)$ \\
$\mathbb{P}^{6} \times \mathbb{P}^{2} $ & $8470728\,\, \zeta(7)$ \\
$\mathbb{P}^{6} \times \mathbb{P}^{1} \times \mathbb{P}^{1} $ & $7529536\,\, \zeta(7)$ \\ \hline
$\mathbb{P}^{9} $ & $263973600\,\, \zeta(3)\,\, \zeta(5)$ \\
$\mathbb{P}^{8} \times \mathbb{P}^{1} $ & $204909696\,\, \zeta(3)\,\, \zeta(5)$ \\
$\mathbb{P}^{7} \times \mathbb{P}^{2} $ & $180006912\,\, \zeta(3)\,\, \zeta(5)$ \\
$\mathbb{P}^{7} \times \mathbb{P}^{1} \times \mathbb{P}^{1} $ & $160006144\,\, \zeta(3)\,\, \zeta(5)$ \\
$\mathbb{P}^{6} \times \mathbb{P}^{3} $ & $167713280\,\, \zeta(3)\,\, \zeta(5)$ \\
$\mathbb{P}^{6} \times \mathbb{P}^{2} \times \mathbb{P}^{1} $ & $141613920\,\, \zeta(3)\,\, \zeta(5)$ \\
$\mathbb{P}^{6} \times \mathbb{P}^{1} \times \mathbb{P}^{1} \times \mathbb{P}^{1} $ & $125879040\,\, \zeta(3)\,\, \zeta(5)$ \\ \hline
$\mathbb{P}^{10} $ & $3748096000/3\,\, \zeta(3)^3 + 17291616320/3\,\, \zeta(9)$ \\
$\mathbb{P}^{9} \times \mathbb{P}^{1} $ & $967032000\,\, \zeta(3)^3 + 4444444440\,\, \zeta(9)$ \\
$\mathbb{P}^{8} \times \mathbb{P}^{2} $ & $846106560\,\, \zeta(3)^3 + 3874204890\,\, \zeta(9)$ \\
$\mathbb{P}^{8} \times \mathbb{P}^{1} \times \mathbb{P}^{1} $ & $752094720\,\, \zeta(3)^3 + 3443737680\,\, \zeta(9)$ \\
$\mathbb{P}^{7} \times \mathbb{P}^{3} $ & $2349916160/3\,\, \zeta(3)^3 + 10737418240/3\,\, \zeta(9)$ \\
$\mathbb{P}^{7} \times \mathbb{P}^{2} \times \mathbb{P}^{1} $ & $661929984\,\, \zeta(3)^3 + 3019898880\,\, \zeta(9)$ \\
$\mathbb{P}^{7} \times \mathbb{P}^{1} \times \mathbb{P}^{1} \times \mathbb{P}^{1} $ & $588382208\,\, \zeta(3)^3 + 2684354560\,\, \zeta(9)$ \\
$\mathbb{P}^{6} \times \mathbb{P}^{4} $ & $2255352400/3\,\, \zeta(3)^3 + 10294287500/3\,\, \zeta(9)$ \\ \hline
  \end{tabular}
\caption{Perturbative corrections $\chi_n$ to the nonlinear $\sigma$-models with Calabi--Yau hypersurface target spaces embedded in the products of projective spaces $\mathbb{P}^{n+1}_\otimes$.}
\label{table:evidence}
\end{table}

In section~\ref{sec:pert-NLSM} we argued that the perturbative
corrections to the metric on the K\" ahler moduli space of two-dimensional $\mathcal{N}=(2,2)$ 
nonlinear $\sigma$-models with $n$-dimensional Calabi--Yau target spaces $X$ are determined by universal
polynomials $\chi_k$ in Chern classes up to degree $n$. We further argued in section~\ref{sec:proposal} that these
polynomials are calculated by the two-sphere partition function according to eq.~\eqref{the_answer}.

In the derivation in section~\ref{sec:proposal} we have assumed that the considered class of Calabi--Yau manifolds is sufficiently generic, i.e., that the Chern monomials are sufficiently distinct such that an unambiguous answer for the universal polynomials $\chi_k$ can be derived. By studying explicit examples we demonstrates here that up to Calabi--Yau ninefolds this assumption is indeed justified.

For this purpose it suffices to analyze the top-degree polynomial $\chi_n$ of Calabi--Yau $n$-folds, as the lower-degree polynomials $\chi_k$ (for $k<n$) have inductively already been confirmed as the top-degree polynomial of Calabi--Yau $k$-folds. Thus, by studying explicit examples we confirm in this appendix that for Calabi--Yau $n$-folds the universal relation holds
\begin{equation}
  \int_X \chi_n \,=\, \left. Z_{S^2}^{\rm pert} \right|_{t=0} \ ,
\label{eqn:match}
\end{equation}
where $\chi_n$ is determined from the $\widehat{\Gamma}^{\phantom{.}\mathbb C}_X$-class or from the $\Lambda_X$-class as discussed in section~\ref{sec:proposal}. We will do this
by computing enough examples to ensure that $\chi_n$ is the only polynomial in Chern classes that could have given the same result as computed by the partition function~$Z_{S^2}$.
In this way, we explicitly verify the universal classes  up through Calabi--Yau ninefolds as given in \eqref{eq:first-few}.

For a Calabi--Yau $n$-fold there are $p(n)-p(n-1)$ non-trivial Chern monomials
of degree $n$, requiring at least as many examples. For $n=\{3,4,5,6,7,8,9\}$
the number of degree $n$ Chern monomials are $\{1,2,2,4,4,7,8\}$. (As an example,
the Chern monomials for $n=7$ are $c_7,c_5c_2,c_4c_3$ and $c_3c_2c_2$.)
Thus, we must compute $28$ examples in all. 

Consider Calabi--Yau hypersurfaces in products of projective spaces
$\mathbb{P}^{n+1}_\otimes=\mathbb{P}^{n_1} \times \dots \times \mathbb{P}^{n_s}$
with $n + 1\equiv \sum n_s$.  As argued in section~\ref{sec:proposal}, there are
$p(n+1)>p(n)-p(n-1)$ such hypersurfaces, and therefore this class
seems to be sufficient for our purposes. In table~\ref{table:evidence}
we present our results. For the considered Calabi-Yau manifolds at dimension~$n$, the
$(p(n)-p(n-1))\times (p(n)-p(n-1))$ matrix, whose rows are the degree $n$
Chern monomials for the Calabi-Yau $n$-fold hypersurfaces in the
ambient spaces $\mathbb{P}^{n+1}_\otimes$, has full rank. Therefore, if the match in equation
\ref{eqn:match} holds for each $n$-fold example, this full rank
condition is sufficient to ensure that the universal polynomial argued
for in section \ref{sec:pert-NLSM} must be $\chi_n$. We have computed
both sides of eq.~\ref{eqn:match} and in all cases we find agreement. For the
consider Calabi--Yau hypersurfaces the resulting perturbative corrections $\chi_n$
are list in table \ref{table:evidence}.

\clearpage

\def\cprime{$'$}

\ifx\undefined\bysame
\newcommand{\bysame}{\leavevmode\hbox to3em{\hrulefill}\,}
\fi

\end{document}